\documentclass[12pt]{article}
\usepackage{cite}
\usepackage{epsfig}
\usepackage{palatino}
\usepackage{graphicx}
\usepackage{amssymb}
\advance\textheight by \topskip
\newcommand{\sm}{\rm SM}
\newcommand{\bz}{\cal{BZ}}

\begin{document}
\tolerance=100000
\thispagestyle{empty}
\setcounter{page}{0}


\baselineskip 16.6pt

\begin{flushright}
SINP/TNP/2007-15\\
\end{flushright}
\vskip 65pt
\begin{center}
{\Large \bf Unparticle physics at hadron collider via dilepton production}\\
\vspace{8mm}
{\bf
Prakash Mathews$^a$
\footnote{prakash.mathews@saha.ac.in}, 
V. Ravindran$^b$
\footnote{ravindra@mri.ernet.in}
}\\
\end{center}
\vspace{10pt}
\begin{flushleft}
{\it
a) 
Saha Institute of Nuclear Physics, 1/AF Bidhan Nagar,
Kolkata 700 064, India.\\

b)
Regional Centre for Accelerator-based Particle Physics,\\ 
Harish-Chandra Research Institute,
Chhatnag Road, Jhunsi, Allahabad, India.

}
\end{flushleft}

\vspace{10pt}
\begin{center}
{\bf ABSTRACT}
\end{center}
\vskip12pt

The scale invariant unparticle physics recently proposed by Georgi could 
manifest at low energies as non integral number $d_{\cal U}$ of invisible 
particles.  Unparticles if existing, could couple to the Standard Model 
fields and consequently affect the collider phenomenology.  We consider 
the DY process to explore effects of the peculiar propagator of the scalar 
and tensor unparticle operators.  To probe these effects at hadron collider 
one needs to go beyond LO in QCD and hence the quantitative impact of QCD 
corrections for unparticle physics at LHC is investigated.  We present the 
K-factors at LHC.  Inclusion of QCD corrections to NLO stabilises the cross 
section with respect to scale variations.  

\vfill
\clearpage

\setcounter{page}{1}
\pagestyle{plain}

\section{Introduction}

Banks and Zaks \cite{bz} found a non-trivial zero of the $\beta$ function
in the IR region of Yang-Mills theories with certain non-integral number 
of fermions, implying absence of a particle like interpretation.  Georgi 
in a recent paper \cite{hep-ph/0703260}, termed this scale invariant stuff as 
unparticle and formulated a frame work in which one could address its 
phenomenological consequences.  The scheme proposed is that at high 
energies the Standard Model ($\sm$) and the Banks-Zaks ($\bz$) fields 
interact via the exchange of particles of mass $M_{\cal U}$
\begin{eqnarray}
\frac{1}{M_{\cal U}^k} O_{\sm} O_{\bz} ~,
\label{eq1}
\end{eqnarray}
where $O_{\sm}$ is the $\sm$ operators of mass dimension $d_{\sm}$ and
$O_{\bz}$ is the $\bz$ operator of mass dimension $d_{\bz}$.  At low 
energy what is relevant is the remanent unparticle and its interaction
with the $\sm$ which is described in the effective field theory.  With
the onset of scale invariance in the $\bz$ sector at some scale 
$\Lambda_{\cal U}$, renormalisation effects induce dimensional 
transmutation.  Below this scale, $\bz$ operators match onto unparticle
operators leading to a new set of interactions
\begin{eqnarray}
C_{\cal U} \frac{\Lambda_{\cal U}^{d_{\bz} - d_{\cal U}}}{M^k_{\cal U}} O_{\sm}
O_{\cal U} ~,
\label{eq2}
\end{eqnarray}
where $C_{\cal U}$ is a coefficient in the low energy effective theory and 
$O_{\cal U}$ the unparticle operator with scaling dimension $d_{\cal U}$.
Further $M_{\cal U}$ should be large enough such that its coupling to $\sm$
must be sufficiently weak, consistent with current experimental data.  In 
general the effective coupling in Eq.~(\ref{eq2}) of the $\sm$ to the 
unparticle could be 
\begin{eqnarray}
\frac{\lambda_S}{\Lambda^{d_{\cal U}}_{\cal U}} T_{\mu}^\mu ~O_{\cal U} ,
\qquad
\frac{\lambda_V}{\Lambda^{d_{\cal U}-1}_{\cal U}} \bar{\psi} \gamma_\mu \psi 
~O^\mu_{\cal U} ,
\qquad
\frac{\lambda_T}{\Lambda^{d_{\cal U}}_{\cal U}} T_{\mu\nu} ~O^{\mu\nu}
_{\cal U} ~.
\label{eq3}
\end{eqnarray}
The dimensionless coupling $\lambda_{\kappa}$ corresponds to the
unparticle operator $O^\kappa_{\cal U}$, where $\kappa=S,V,T$
refers to the scalar, vector and tensor operators respectively.  
$T_{\mu\nu}$ is the energy momentum tensor of the $\sm$ and 
$d_{\cal U}$ the scaling dimension of the unparticle operators 
$O^\kappa_{\cal U}$.  These operators are Hermitian and $O^\mu_{\cal U}$ 
and $O^{\mu\nu}_{\cal U}$ are transverse.  Given these effective 
interactions Eq.~(\ref{eq3}), unparticle stuff could be produced 
by a single insertion of the interaction from a $\sm$ 
process leading to a missing energy and momentum signals.  The phase 
space for the unparticle production corresponding to the operators 
$O_{\cal U}$ of scaling dimension $d_{\cal U}$ is the same as the 
production of $d_{\cal U}$ (non-integral number) invisible 
particles and is proportional to the factor \cite{hep-ph/0703260}
\begin{eqnarray}
A_{d_{\cal U}}=\frac{16 \pi^{5/2}}{(2 \pi)^{2 d_{\cal U}}}
\frac{\Gamma(d_{\cal U}+1/2)}{\Gamma(d_{\cal U}-1)\Gamma(2 d_{\cal U})}~~.
\nonumber
\end{eqnarray}
Further the exchange of a virtual unparticle corresponding to an operator
$O^{\kappa}_{\cal U}$ between the $\sm$ particles, would need the propagator
for the unparticle.  Using scale invariance and transverse properties
of the vector and tensor operators $O^{\kappa}_{\cal U}$, the unparticle 
propagator was obtained \cite{hg2,kingman}
\begin{eqnarray}
\int d^4 x e^{i P \cdot x} 
<0|T O^{\kappa}_{\cal U} (x) O^{\kappa}_{\cal U} (0)|0>
&=&\frac{i A_{d_{\cal U}}}{2 \sin(d_{\cal U} \pi)} \frac{B_{\kappa}}
{(-P^2-i \epsilon)^{2-d_{\cal U}}} ~,
\label{eq4}
\end{eqnarray}
where $B_{\kappa}$ depends on the Lorentz structure of the operator 
$O_{\cal U}$ as given below:
\begin{center}
\begin{tabular}{ccc}
$O_{\cal U}$ & \qquad \qquad &1
\\[2ex]
$O^\rho_{\cal U}$ & & $\eta_{\mu\nu}(P)= -g_{\mu\nu}+ \frac{P_\mu P_\nu}{P^2}$
\\[2ex]
$O^{\rho \sigma}_{\cal U}$ & & $B_{\mu\nu\alpha\beta}(P)={1 \over 2}
\left(\eta_{\mu\alpha} 
\eta_{\nu\beta}+\eta_{\mu\beta} \eta_{\nu\alpha}- 
\frac{2}{n-1}\eta_{\mu\nu} \eta_{\alpha\beta}\right)$ ~~,
\end{tabular}
\end{center}
where, $n$ number of space time dimension.
The propagator of the unparticle is singular due to the $\sin (d_{\cal U} 
\pi)$ factor in the denominator and hence the $d_{\cal U}$ is constrained 
in the range $1<d_{\cal U}<2$.  The phenomenological consequences of 
unparticles are dictated by the interactions given in Eq.~(\ref{eq3}),
the nature of the unparticle phase space and the propagator.  Given 
these sufficient informations, we can calculate various quantities of 
phenomenological interest, without actually having to understand the 
nature of the unparticle {\em per se}.  Close to the scale 
$\Lambda_{\cal U}$, say of the order of a TeV the unparticle
effects could manifest and hence be relevant at the LHC and ILC.  There 
has been some activity to explore possible collider and flavour
phenomenology of unparticle physics \cite{hg2}-\cite{chen-he}.  Unparticle
effects on cosmology and astrophysics have been considered in \cite{hooman}.

In this paper we consider the Drell-Yan (DY) process at LHC and study the 
effects of unparticle corresponding to scalar and tensor operators which 
could also produce the dilepton pair.  The contribution from the vector 
unparticle operator was considered in \cite{kingman} for the DY process.  
At hadron colliders, a precise measurement of dilepton production cross 
sections is possible and hence could be sensitive to the physics beyond 
the $\sm$ through the exchange of unparticles.  At hadron colliders such 
as Tevatron and LHC, the theoretical uncertainties coming from QCD effects 
due to initial state partons are quite sizable.  The sources of these 
uncertainties are the renormalisation and factorisation scale dependence.  
Experience with next-to-leading order (NLO) QCD contributions to SM processes 
{\it viz.} DY process and Higgs production at hadron colliders strongly 
suggests that the leading order (LO) results are quite unreliable.  The 
NLO contributions are large and in addition the theoretical uncertainties 
are significantly reduced.  To make any quantitative statement about 
unparticle searches at hadron colliders it is necessary to look at the 
NLO-QCD corrections to the DY process.  In this analysis we have studied 
the effects of scalar and tensor unparticle propagator on the invariant 
mass distribution of the dilepton.

\section{Drell-Yan process}

Consider the production of dileptons $\ell^+$ and $\ell^-$ in the collision 
of hadrons $P_1$ and $P_2$
\begin{eqnarray}
P_1(p_1)+P_2(p_2) \rightarrow \ell^+(l_1)+\ell^-(l_2)+X(P_X)  ~,
\label{eq5}
\end{eqnarray}
where the final inclusive hadronic state is denoted by $X$ and it carries a
momentum $P_X$.  In the QCD improved parton model, the hadronic cross section
can be expressed in terms of partonic cross sections $d \hat \sigma^{ab}$ 
convoluted with appropriate parton (of type $a$) distribution functions 
$f_a^{P}(x)$ as follows
\begin{eqnarray}
2 S~{d \sigma^{P_1 P_2} \over d Q^2}\left(\tau,Q^2\right)
=\sum_{ab={q,\overline q,g}} \int_0^1 dx_1 ~ dx_2~ dz ~
f_a^{P_1}(x_1) ~
f_b^{P_2}(x_2)
\,\, 
2 \hat s ~
{d \hat \sigma^{ab} \over d Q^2}\left(z,Q^2\right)
\delta(\tau-z x_1 x_2)\,.
\label{eq6}
\end{eqnarray}
The scaling variables are defined by $k_1 =x_1 p_1$, $ k_2=x_2 p_2$, where 
$k_1,k_2$ are the momenta of incoming partons,
\begin{eqnarray}
(p_1+p_2)^2 &\equiv& S, \quad \quad \quad
(k_1+k_2)^2 \equiv \hat s, \quad \quad \quad (l_1+l_2)^2=q.q \equiv Q^2,
\nonumber\\[2ex]
\tau&=&{Q^2 \over S},
\quad \quad \quad
z={Q^2 \over \hat s },
\quad \quad \quad \tau=x_1 x_2 z.
\label{eq7}
\end{eqnarray}
The dileptons can be produced through the partonic cross sections given by 
$a(k_1)+b(k_2) \rightarrow j(-q) +\displaystyle \sum_i^m X_i(-p_i)$, where 
$j$ could be the usual $\sm$ photon, $Z$-boson or an unparticle stuff ${\cal
U}$ which would decay to a leptonic pair in the final state.  
At LO in QCD, the contribution from the $\sm$ is only through 
the quark initiated process $q + \bar q \rightarrow \gamma^*/Z^* \rightarrow
\ell^+ + \ell^-$.  When 
unparticle fields couple to SM fields, they 
could also decay to a pair of leptons through quark initiated process 
$ q + \bar q \rightarrow {\cal U} \rightarrow \ell^+ + \ell^-$ 
and also through gluon initiated process 
$ g + g \rightarrow {\cal U} \rightarrow \ell^+ + \ell^-$.  
As we indicated in the previous section, if the unparticle fields couple to 
the SM fields through vector interaction,
then the LO hadronic cross section can get contribution 
only from quark, anti-quark
initiated process.  On the other hand, if the interaction of the unparticle
fields with the $\sm$ is of scalar
or tensorial in nature, then gluons can also produce unparticle fields  
and hence contribute to LO in QCD.  Gluon initiated processes
at LHC often give large contributions because its flux is the largest compared
to other partonic fluxes that contribute to various cross sections.    
Since the higher order QCD effects
can stabilise the cross section, we have incorporated all the  
NLO-QCD effects through the following processes
\begin{eqnarray}
&&q + \bar q \rightarrow \gamma^*/Z^* + g\,, \quad \quad q+\bar q
\rightarrow \gamma^*/Z^* + \mbox{one~~loop}\,,
\nonumber \\[2ex]
&&q + g \rightarrow \gamma^*/Z^* + q\,,
\qquad \bar q + g \rightarrow \gamma^*/Z^* + \bar q\,,
\label{eq8}
\end{eqnarray}
in the $\sm$ and in unparticle theories,
\begin{eqnarray}
&&q + \bar q \rightarrow {\cal U} + g\,, \quad \quad q+\bar q \rightarrow 
{\cal U} +
\mbox{one~~loop}\,,
\nonumber \\[2ex]
&&q + g \rightarrow {\cal U}  + q\,,
\qquad \bar q + g \rightarrow {\cal U}  + \bar q\,,
\nonumber \\[2ex]
&&g + g \rightarrow {\cal U} + g \,, \quad \quad g+ g \rightarrow {\cal U}+
\mbox{one~~loop}\,.
\label{eq9}
\end{eqnarray}
In the above, we consider all the one loop QCD corrections to
quark antiquark unparticle vertex and gluon gluon unparticle vertex.

The cross sections beyond LO involve the computation of one loop virtual 
corrections and real bremsstrahlung contributions to LO processes.  The 
singularities we encounter in our computation are due to the soft and 
collinear divergences.  We have used the dimensional regularisation to
regulate both these singularities.  The soft divergence coming from the
virtual gluons and the bremsstrahlung contributions cancel exactly according
to the Bloch-Nordsieck theorem.  The remaining collinear divergences are 
removed by mass factorisation for which we use the $\overline {MS}$ scheme.
Hence the complete cross section after mass factorisation is
\begin{eqnarray}
2 S~{d \sigma^{P_1P_2} \over dQ^2}(\tau,Q^2)&=&
\sum_q{\cal F}_{SM,q} \int_0^1~ {d x_1}~ \int_0^1
~{dx_2}~ \int_0^1~ dz~ \delta(\tau-z x_1 x_2)
\nonumber\\[2ex]&&
\times \Bigg[ H_{q \bar q}(x_1,x_2,\mu_F^2) \Big(
\Delta_{q \bar q}^{(0),\gamma/Z}(z,Q^2,\mu_F^2)
 +a_s \Delta_{q \bar q}^{(1),\gamma/Z}(z,Q^2,\mu_F^2)\Big)
\nonumber\\[2ex] &&
+\Big( H_{q g}(x_1,x_2,\mu_F^2)+H_{g q}(x_1,x_2,\mu_F^2)\Big)
 a_s \Delta_{q g}^{(1),\gamma/Z}(z,\mu_F^2) \Bigg]
\nonumber\\[2ex]
&&+\sum_q{\cal F}_{\cal U} \int_0^1~ {d x_1 }~ \int_0^1
~{dx_2}~ \int_0^1~ dz~ \delta(\tau-z x_1 x_2)
\nonumber\\[2ex]&&
\times \Bigg[ H_{q \bar q}(x_1,x_2,\mu_F^2) \Big(
\Delta_{q \bar q}^{(0),{\cal U}}(z,Q^2,\mu_F^2)
 +a_s \Delta_{q \bar q}^{(1),{\cal U}}(z,Q^2,\mu_F^2)\Big)
\nonumber\\[2ex] &&
+\Big( H_{q g}(x_1,x_2,\mu_F^2)+H_{g q}(x_1,x_2,\mu_F^2)\Big)
 a_s \Delta_{q g}^{(1),{\cal U}}(z,Q^2,\mu_F^2)
\nonumber\\[2ex]&&
+ H_{g g}(x_1,x_2,\mu_F^2) \Big(
\Delta_{g g}^{(0),{\cal U}}(z,Q^2,\mu_F^2)
 +a_s \Delta_{g g}^{(1),{\cal U}}(z,Q^2,\mu_F^2)\Big) \Bigg]\, ,
\label{eq10}
\end{eqnarray}
where $\Delta^{(i) j} (z,Q^2,\mu_F^2), j=\gamma^*/Z^*, {\cal U}$ 
are the mass factorised coefficient functions of the $\sm$ and 
unparticle theories respectively to order $i$ in QCD.
The constants ${\cal F}_{SM,q}$ and ${\cal F}_{G}$ are given by
\begin{eqnarray}
{\cal F}_{SM,q}&=&{4 \alpha^2 \over 3 Q^2} \Bigg[Q_q^2
-{2 Q^2 (Q^2-M_Z^2) \over
\left((Q^2-M_Z^2)^2+M_Z^2 \Gamma_Z^2\right) c_w^2 s_w^2}
Q_q g_e^V g_q^V
\nonumber\\[2ex]
&&+{Q^4 \over  \left((Q^2-M_Z^2)^2+M_Z^2 \Gamma_Z^2\right)
c_w^4 s_w^4}\Big((g_e^V)^2+(g_e^A)^2\Big)\Big((g_q^V)^2+(g_q^A)^2\Big)
\Bigg]\,,
\nonumber\\[2ex]
{\cal F}^\kappa_{{\cal U}}&=& {
{\cal C}_\kappa 
\lambda_\kappa^4 
A^2_{d_{\cal U}}
\over 
4 \pi^2 
\sin^2(\pi d_{\cal U}) Q^2} 
\left( Q^2 \over 
\Lambda_{{\cal U}}^2\right)^{2 d_{\cal U}-m} ~,
\label{eq11}
\end{eqnarray}
where $\lambda_\kappa$ is varied from $0.4$ to $0.9$ for 
$\kappa=$ scalar (S), tensor (T).  For the tensorial interaction, 
the constant ${\cal C}_{T}=1/80$ and $m=0$,
for the scalar interaction ${\cal C}_{S}=1/32 $ for both quark anti-quark
($m=2$) as well as gluon ($m=1$) initiated processes.  

The renormalised partonic distributions are
\begin{eqnarray}
H_{q \bar q}(x_1,x_2,\mu_F^2)&=&
f_q^{P_1}(x_1,\mu_F^2)~
f_{\bar q}^{P_2}(x_2,\mu_F^2)
+f_{\bar q}^{P_1}(x_1,\mu_F^2)~
f_q^{P_2}(x_2,\mu_F^2)\,,
\nonumber
\\[2ex]
H_{g q}(x_1,x_2,\mu_F^2)&=&
f_g^{P_1}(x_1,\mu_F^2) ~
\Big(f_q^{P_2}(x_2,\mu_F^2)
+f_{\bar q}^{P_2}(x_2,\mu_F^2)\Big)\,,
\nonumber
\\[2ex]
H_{q g}(x_1,x_2,\mu_F^2)&=&
H_{g q}(x_2,x_1,\mu_F^2)\,,
\nonumber
\\[2ex]
H_{g g}(x_1,x_2,\mu_F^2)&=&
f_g^{P_1}(x_1,\mu_F^2)~
f_g^{P_2}(x_2,\mu_F^2)\,.
\label{eq12}
\end{eqnarray}

\section{Discussions}
In this section, we present the impact of unparticle fields at NLO level 
in QCD to the dilepton production at LHC with the center of mass energy 
$\sqrt{S}=14$ TeV.  Similar effects can be easily studied at Tevatron 
using our results and we postpone the detailed study for future publication.  
We have not used the existing Tevatron data
to constrain the parameters of unparticle theories, instead we
have demonstrated the observable effects of this new theory
using reasonable choices of parameters.  Our predictions 
are less sensitive to renormalisation and factorisation scale 
uncertainties.  The $\sm$ parameters are $\alpha=1/137.03604$,
$M_Z=91.1876$ GeV, $\Gamma_Z=2.4952$ GeV.  For the parton density sets,
we have used in the leading order the MRST 2001 LO ($\Lambda=0.1670$ GeV) and
in the next-to-leading order the MRST 2001 NLO ($\Lambda=0.2390$ GeV).  
Recent MRST sets available in the literature are not expected to change our findings.

\begin{figure}[!h]
\begin{center}
\epsfig{file=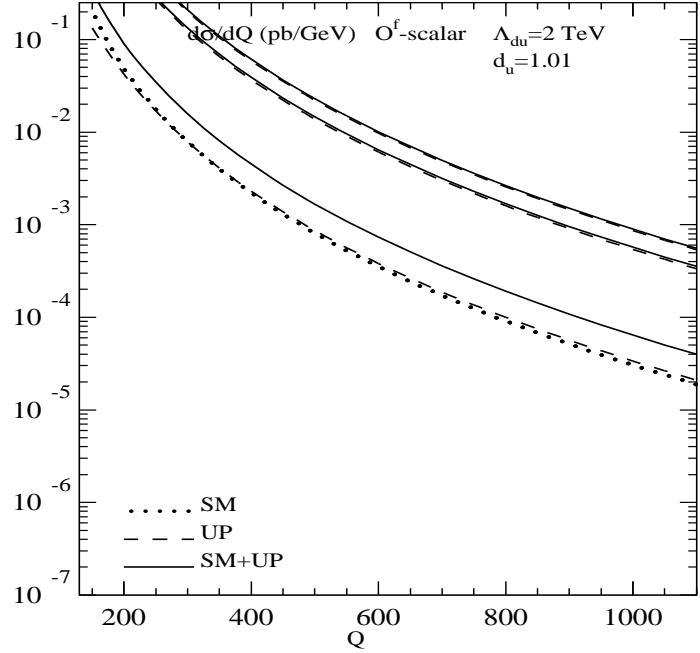,width=9.7cm,height=9.7cm,angle=0}
\vspace*{-0.2cm}
\caption{\em Invariant mass distribution of the 
dilepton which includes the scalar unparticles coupling to
quark-antiquark.  The lowest set corresponds to $\lambda_S=0.4$,
middle $\lambda_S=0.8$ and upper $\lambda_S=0.9$. 
}
\label{fig1}
\end{center}
\end{figure}

\begin{figure}[!h]
\begin{center}
\epsfig{file=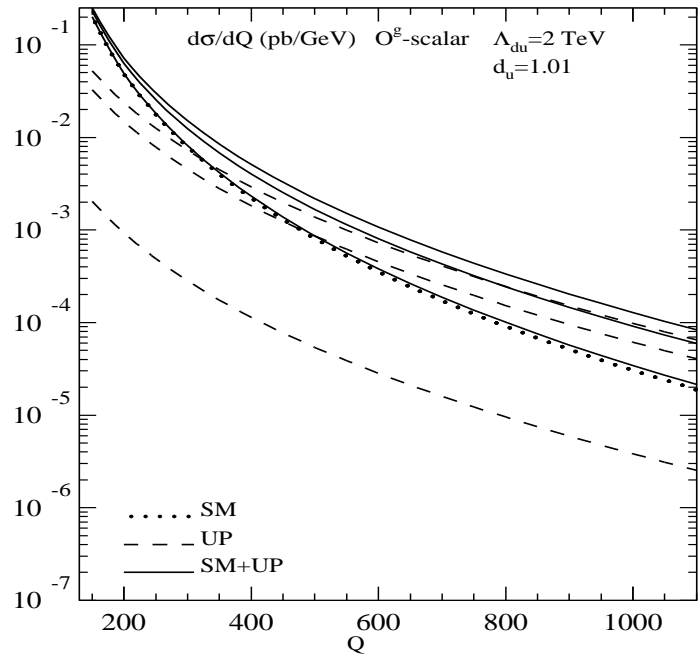,width=9.7cm,height=9.7cm,angle=0}
\vspace*{-0.2cm}
\caption{\em The scalar unparticles coupling to gluons
in the initial state to produce a dilepton pair with
invariant mass $Q$.  
The lowest set corresponds to $\lambda_S=0.4$,
middle $\lambda_S=0.8$ and upper $\lambda_S=0.9$. 
}
\label{fig2}
\end{center}
\end{figure}

\begin{figure}[!h]
\begin{center}
\epsfig{file=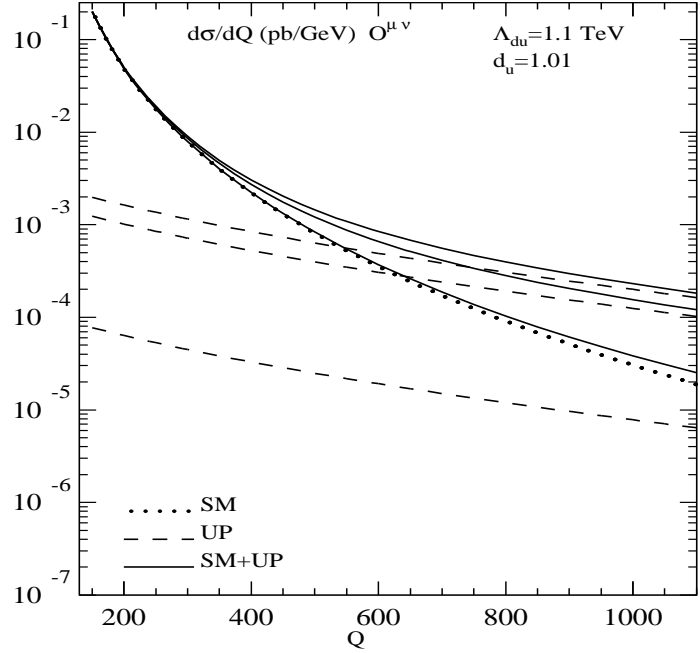,width=9.7cm,height=9.7cm,angle=0}
\vspace*{-0.2cm}
\caption{\em The tensor unparticles coupling to quarks and gluons
in the initial state to produce a dilepton pair with
invariant mass $Q$.  
The lowest set corresponds to $\lambda_T=0.4$,
middle $\lambda_T=0.8$ and upper $\lambda_T=0.9$. 
}
\label{fig3}
\end{center}
\end{figure}

The invariant mass distributions for scalar (both quark-antiquark and gluon
initiated processes) and tensor interactions are plotted in the 
Fig.~(\ref{fig1}-\ref{fig3}) for the range $150 ~{\rm GeV} < Q< 1100$ 
GeV and  for the coupling parameters $\lambda_k=0.4$ (lowest), $0.8$ (middle), 
$0.9$ (upper).  We choose 
$d_{\cal U}=1.01$ for all the plots.   It is clear from
the Figs..~(\ref{fig1},\ref{fig2},\ref{fig3}), the unparticle effects will be 
visible only for larger values of $\lambda_k$.  
We also find that the unparticle fields tend to increase the cross section in the large invariant 
mass region, due to the factor $(Q^2)^{d_{\cal U}}$ in 
${\cal F}^\kappa_{\cal U}$.  
This effect is very similar to the four Fermi interaction or $Z'$ 
exchange or tower of Kaluza-Klein excitations in extra-dimensional models
(see \cite{us}).  The nature of these interactions are similar but for the 
overall scale dependent couplings.

\begin{figure}[!h]
\begin{center}
\epsfig{file=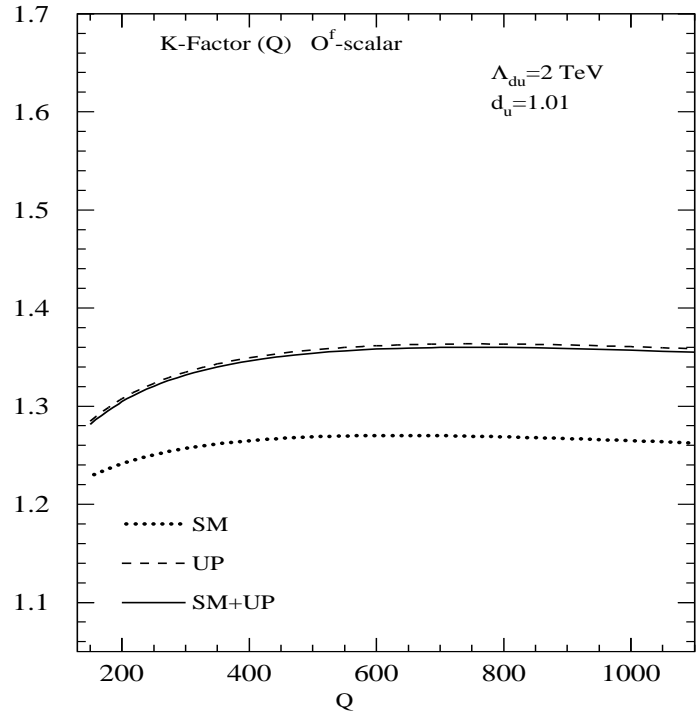,width=9.7cm,height=9.7cm,angle=0}
\vspace*{-0.2cm}
\caption{\em $K$ factor for the 
invariant mass distribution of dileptons at LO and NLO for the 
scalar unparticle from the quark-antiquark channel for $\lambda=0.9$.
}
\label{fig4}
\end{center}
\end{figure}

\begin{figure}[!h]
\begin{center}
\epsfig{file=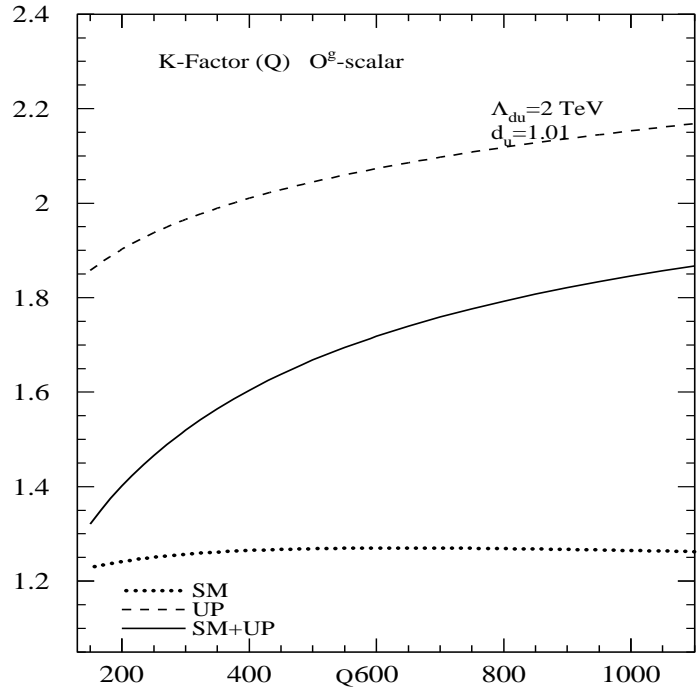,width=9.7cm,height=9.7cm,angle=0}
\vspace*{-0.2cm}
\caption{\em The K-factor for the invariant mass distribution of dilepton 
production in which the scalar unparticle is exchanged from the 
gluon-gluon channel for $\lambda=0.9$.
}
\label{fig5}
\end{center}
\end{figure}

\begin{figure}[!h]
\begin{center}
\epsfig{file=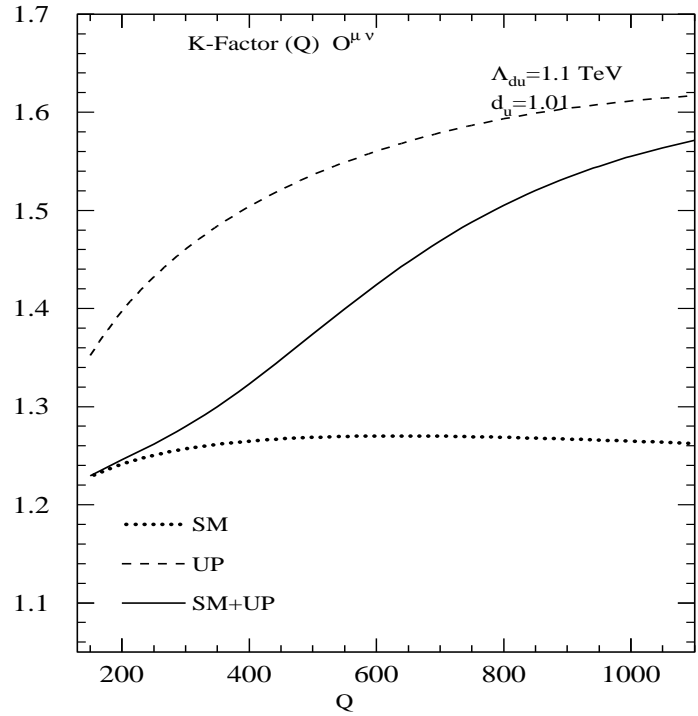,width=9.7cm,height=9.7cm,angle=0}
\vspace*{-0.2cm}
\caption{\em The K-factor for the invariant mass distribution of dilepton 
production in which the tensor unparticle is exchanged from the 
quarks and gluons in the initial state for $\lambda=0.9$.
}
\label{fig6}
\end{center}
\end{figure}

The higher order QCD effect is quantified by the K factor defined by
\begin{eqnarray} 
K=\left[ {d \sigma_{LO}(Q) \over d Q} \right]^{-1} 
   \left[ {d \sigma_{NLO} (Q) \over dQ} \right].
\label{eq13}
\end{eqnarray}
We have plotted the K factor for both scalar and tensor interactions in 
Fig.~(\ref{fig4}-\ref{fig6}). We find that the K factor in the unparticle 
theory for quark anti-quark initiated process is very similar to that of 
SM.  On the other hand, the gluon initiated processes receive large K 
factor for both scalar and tensorial interactions.  

\begin{figure}[!h]
\begin{center}
\epsfig{file=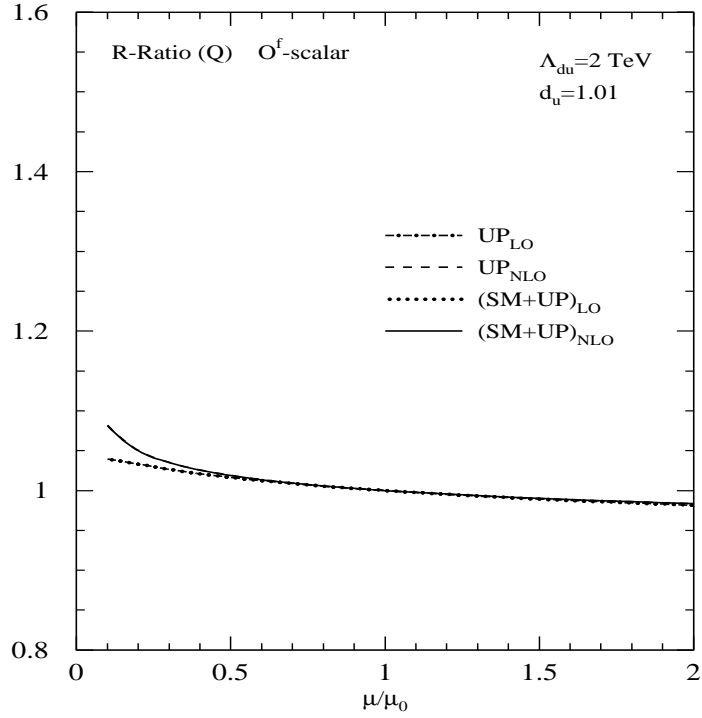,width=9.7cm,height=9.7cm,angle=0}
\vspace*{-0.2cm}
\caption{\em Factorisation scale variation of the 
invariant mass distribution of dileptons at LO and NLO for the 
scalar unparticle from the quark-antiquark channel for $\lambda=0.9$.
}
\label{fig7}
\end{center}
\end{figure}

\begin{figure}[!h]
\begin{center}
\epsfig{file=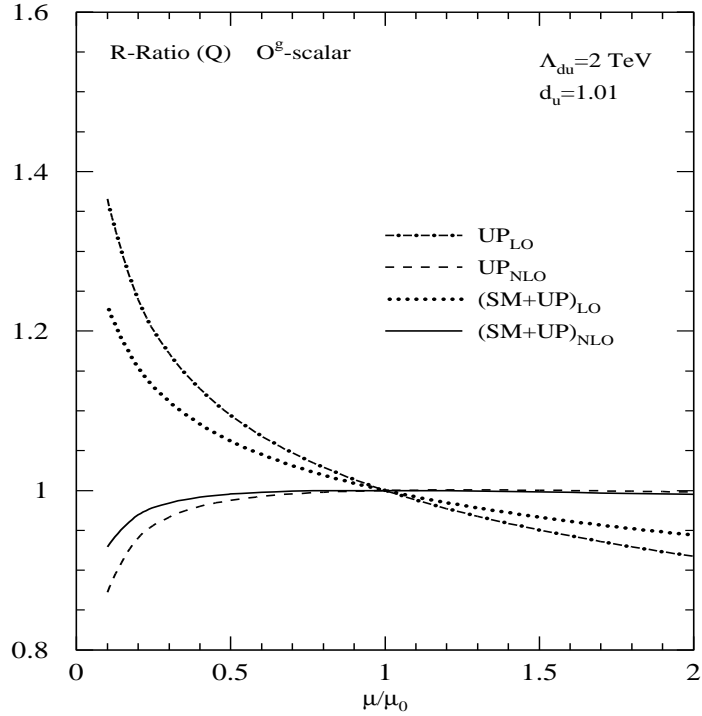,width=9.7cm,height=9.7cm,angle=0}
\vspace*{-0.2cm}
\caption{\em Factorisation scale variation of LO and NLO invariant mass 
distribution of dileptons from the gluon initiated process which couple 
to the scalar unparticle for $\lambda=0.9$.
}
\label{fig8}
\end{center}
\end{figure}

\begin{figure}[!h]
\begin{center}
\epsfig{file=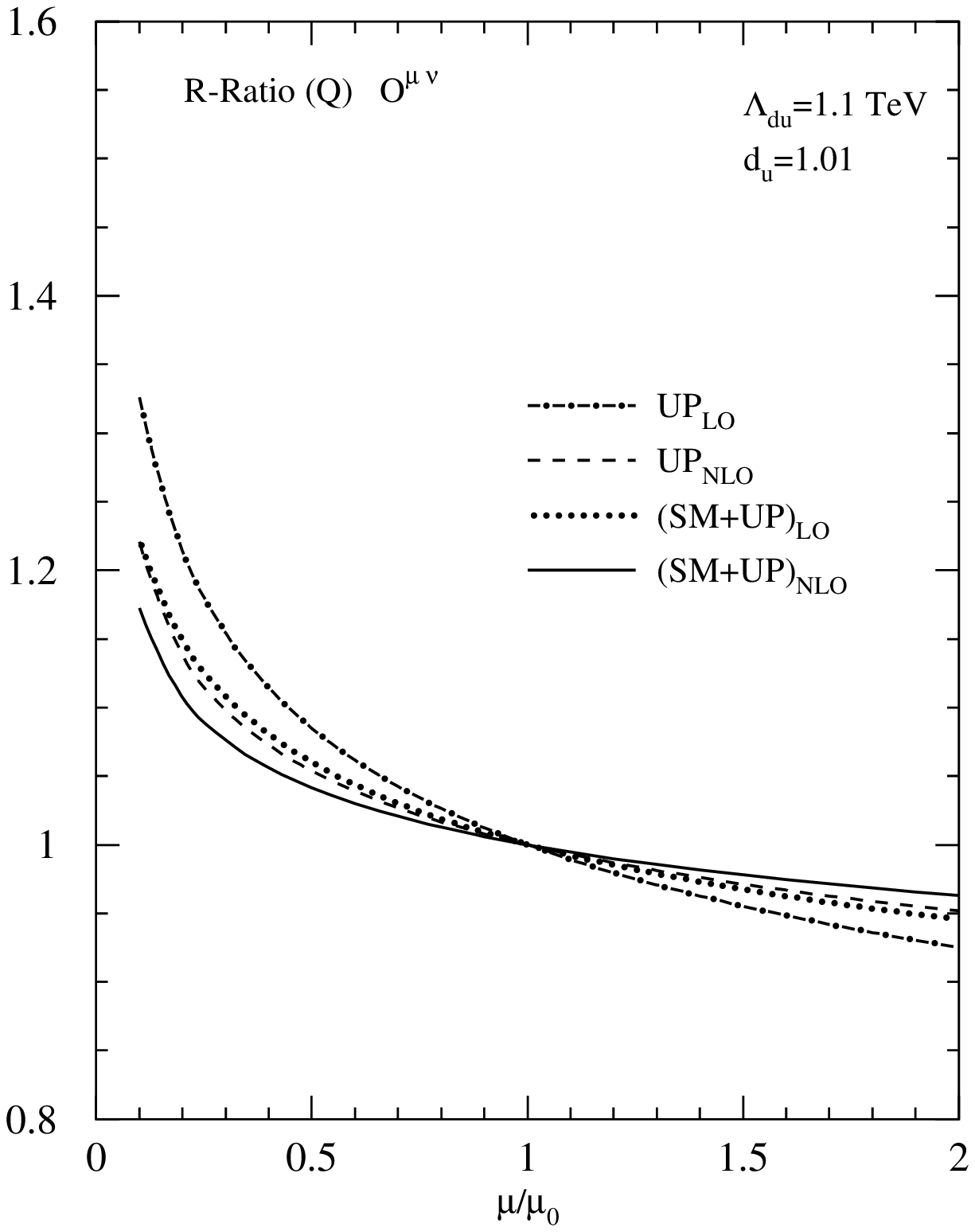,width=9.7cm,height=9.7cm,angle=0}
\vspace*{-0.2cm}
\caption{\em Factorisation scale variation of LO and NLO dilepton invariant 
mass distribution which includes the tensor unparticle.  This includes both 
the quark-antiquark and gluon channels for $\lambda=0.9$.
}
\label{fig9}
\end{center}
\end{figure}

Finally, we have shown how NLO corrections improve the scale uncertainties 
by plotting the following ratios at LO and NLO  
\begin{eqnarray}
R_{LO} &=& \left[ {d \sigma_{LO}(Q,\mu=\mu_0) \over dQ}\right]^{-1}
\left[ {d \sigma_{LO}(Q,\mu) \over dQ} \right] \Bigg|_{Q=700~GeV},
\nonumber \\[2ex]
R_{NLO}&=&\left[ {d\sigma_{NLO}(Q,\mu=\mu_0) \over dQ} \right]^{-1}
\left[ {d \sigma_{NLO}(Q,\mu) \over dQ } \right] \Bigg|_{Q=700~ GeV}
\, .
\label{eq14}
\end{eqnarray}
From the plots in Fig.~(\ref{fig7}-\ref{fig9}), we find that our NLO 
corrected predictions are less sensitive to renormalisation and 
factorisation scales making our predictions stable under QCD radiative 
corrections that are important at hadron colliders.

To summarize our study, we find that both scalar and tensorial
interactions of unparticle fields to $\sm$ fields can lead
to sizable observable effect in the invariant mass distributions of 
of dilepton pairs at hadron colliders in the large invariant mass 
region.  Since QCD effects at hadron colliders are important,
we have incorporated its effect through next-to-leading order 
corrections.  We found that unlike quark initiated processes,
the gluon initiated processes give large K factor which
has to be taken into account when extracting the parameters of the
unparticle theory from the experimental data.  
We have also demonstrated how our QCD improved
next-to-leading order results improve the predictions by reducing
the uncertainties coming from the factorisation scale.

\vspace{.5cm}
\noindent
{\bf Acknowledgments}

\noindent
PM would like to thank the Centre for High Energy Physics, Indian Institute 
of Science, for hospitality where part of this work was done.

\eject



\begin{thebibliography}{99}

\bibitem{bz}
  T.~Banks and A.~Zaks,
  Nucl.\ Phys.\  B {\bf 196}, (1982) 189.

\bibitem{hep-ph/0703260}
  H.~Georgi,
  arXiv:hep-ph/0703260.

\bibitem{hg2}
  H.~Georgi,
  arXiv:0704.2457 [hep-ph].

\bibitem{kingman}
  K.~Cheung, W.~Y.~Keung and T.~C.~Yuan,
  arXiv:0704.2588 [hep-ph].

\bibitem{Luo}
  M.~Luo and G.~Zhu,
  arXiv:0704.3532 [hep-ph].

\bibitem{Chen}
  C.~H.~Chen and C.~Q.~Geng,
  arXiv:0705.0689 [hep-ph].

\bibitem{Ding}
  G.~J.~Ding and M.~L.~Yan,
  arXiv:0705.0794 [hep-ph].

\bibitem{Liao}
  Y.~Liao,
  arXiv:0705.0837 [hep-ph].

\bibitem{Aliev}
  T.~M.~Aliev, A.~S.~Cornell and N.~Gaur,
  arXiv:0705.1326 [hep-ph].

\bibitem{Li}
  X.~Q.~Li and Z.~T.~Wei,
  arXiv:0705.1821 [hep-ph].

\bibitem{Lu}
  C.~D.~Lu, W.~Wang and Y.~M.~Wang,
  arXiv:0705.2909 [hep-ph].

\bibitem{Stephanov}
  M.~A.~Stephanov,
  arXiv:0705.3049 [hep-ph].

\bibitem{Fox}
  P.~J.~Fox, A.~Rajaraman and Y.~Shirman,
  arXiv:0705.3092 [hep-ph].

\bibitem{greiner}
N.~Greiner 
arXiv:0705.3518 [hep-ph].

\bibitem{debu}
D.~Choudhury, D.~K.~Ghosh, Mamta 
arXiv:0705.3637 [hep-ph].

\bibitem{chen-he}
Shao-Long Chen, Xiao-Gang He
arXiv:0705.3946 [hep-ph].

\bibitem{hooman}
H.~Davoudiasl 
arXiv:0705.3636 [hep-ph].

\bibitem{us}

Prakash Mathews, V.~Ravindran, K.~Sridhar and W.L.~van Neerven, 
Nucl. Phys. B 713 (2005) 333.

\end{thebibliography}
\end{document}